# TEduChain: A Platform for Crowdsourcing Tertiary Education Fund using Blockchain Technology

Mahmood A. Rashid
Institute for Integrated and Intelligent Systems
Griffith University, Nathan
Queensland, Australia
mahmood.rashid@griffith.edu.au

Divnesh Prasad
School of Computing Info & Maths
Faculty of Science, Technology & Environment
The University of the South Pacific
Suva, Fiji
s11134755@student.usp.ac.fj

Sarvesh Chand
School of Computing Info & Maths
Faculty of Science, Technology & Environment
The University of the South Pacific
Suva, Fiji
s11133165@student.usp.ac.fj

Krishneel Deo
School of Computing Info & Maths
Faculty of Science, Technology & Environment
The University of the South Pacific
Suva, Fiji
s11119715@student.usp.ac.fj

Kunal Singh
School of Computing Info & Maths
Faculty of Science, Technology & Environment
The University of the South Pacific
Suva, Fiji
s11110926@student.usp.ac.fj

Mansour Assaf
School of Engineering & Physics
Faculty of Science, Technology & Environment
The University of the South Pacific
Suva, Fiji
assaf_m@usp.ac.fj

*Abstract*—Blockchain is an emerging technology framework for creating and storing transaction in distributed ledgers with a high degree of security and reliability. In this paper we present a blockchain-based platform to create and store contracts in between students and their higher education sponsors. The sponsorship might be in any form, such as scholarship, donation or loan. The fund will be arranged and managed by a group of competitive agents (Fundraisers) who will hold the distributed ledgers and act as miners in the blockchain network.

## I. INTRODUCTION

The blockchain [1 and 2] is a technology framework originally known as block chain [3 and 4] is a continuously growing records stored in blocks where the blocks are linked using cryptographic hash.

In this work, we propose a platform for funding needy students for their tertiary education based on the concept of blockchain technology [5]. The system exploits the benefits of the blockchain, as a decentralized architecture (Figure 1), offering secure long term investment mechanism, integrity, and transparency, in order to create a globally trusted funding system for poor undergraduate and post-graduate students who wish to continue their higher education.

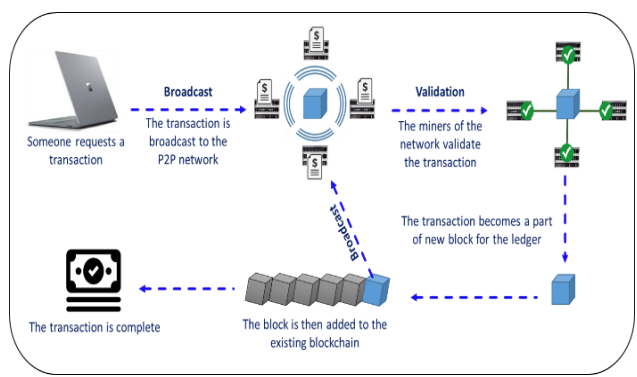

Figure 1 Decentralized Blockchain Architecture

This work was supported in part by the Centre of Flexible Learning (CFL) of the University of the South Pacific (USP), under the TELS project Grant 03072018.

The use of blockchain allows us to track the investors and students knowing that the information is true, as its part of the chain which has been verified due the system being distributed.

It is believed that talented high school students should not be prevented to have the opportunities to pursue university education towards building professional careers that only fortunate students get. By revolutionizing the funding scheme through blockchain technology, we can give every talented student the opportunities that they deserve.

The objectives of the platform can be summarized as follows:

- A blockchain technology based decentralized platform that allows sponsors and students to attain funding by selling future revenue rights to investors worldwide. This could be enforced by directly setting up a policy with the nation's revenue authority to monitor these students, once they start earning a certain amount could be collected from the students' income and sent to the fundraiser who can than reimburse the investors appropriately.
- The investment process is automated and it uses the concept of Contract among student and the sponsors.
- The investments are liquid assets that can be converted to tokens at any time.

The scientific contribution of this work is to provide a distributed and interoperable architecture model for the higher education scholarship, sponsorship, or study loan schemes which addresses global benefits for students, educational institutions and potential investors.

As a proof of concept and for validation purposes, we developed a prototype of the platform from scratch written in Java 8 SE.

The use of blockchain allows us to track the investors and students knowing that the information is true, as its part of the chain which has been verified due the system being distributed. With the traditional system, there is only one instance of the entire record and changing the contract information of any student maliciously would go undetected.

With blockchain, a little tempering with any record would easily be identified.

Students can take advantage of having their university tuition fees and cost of living for the study period covered by a transparent bursary schemes. The investors contributing to this scheme will get reimbursed plus awarded a percentage of what a student would be making from their work after graduation for a predefined period of time. In the following section we present a brief literature review of the technology.

## II. RELATED WORK

Blockchain technology is used in several domains such as healthcare management [7], clinical trials [9], pharmaceutical industry [10], smart business, electronic government [11], sports, logistics [12], intelligent transportation systems [13], and higher education credit systems [6], etc. due to its benefits in distributed data storage, transparency, and decentralized environment where no third party is in control of the transactions [8].

In [14], energy production, management, and trading using smart grids and blockchain technology is discussed and a model for business operation optimization is proposed. The research domain of Internet of Things (IoT) and how it can benefits from blockchain technology including security, management, and privacy issues has been addressed in literature [15, 16, and 17].

Solutions related to higher education using blockchain technology and cryptography such as creating, sharing, and verifying digital academic certificates and credit-based achievements have been investigated and employed by several higher education institutions in Europe and South America [18, 19, and 20].

In [6], a blockchain based global university credit system is presented. They proposed a simplified and efficient solution for sharing student's credit record with other higher education institutions, while eliminating language and administrative obstacles, additionally the system is publicly accessible on the web via internet. The methodology is presented next.

## III. OUR PLATFORM

In this paper, we propose a complete platform named *TEduChain* to collect and manage funds for students for higher study and administer them afterwards. Three different entities – fund seeking students, fundraisers, and sponsors are the actors of the platform. The platform provides customized dashboards for different entities. The entities interact with each other efficiently using their own dashboards. A brief description of the entities are presented below:

- *Students:* Students who are seeking fund for their tertiary study in terms of sponsorship, such as scholarship, donation, and study loan. If the fund comes through the investors as a loan, the students need to repay the loan based on the defined terms and conditions of the loan. The students must express their interest first through the platform to include them into the blockchain network.

- *Fundraisers:* Fundraisers are the licensed business entities who work as intermediary agents to collect funds for the students and after securing the funds they manage the students 'portfolios. The fundraisers are part of the system to as the representative of the students to invite the investors and collect funds for the students. The fundraisers also act as the miners in the network and once the full amount needed by a student is collected, prepare a document (contract) that binds the student and the respective investors. The fundraiser entities are the traditional scholarship bodies located in different countries all over the world. Fundraisers play the important role for validating, authenticating, and maintaining student applications and records.

- *Sponsors:* The sponsors could be individuals or business organizations—scholarship providers, philanthropists, or investors—who are committed to fund for tertiary education through one or multiple fundraisers. Every sponsor has their own terms and conditions for the sponsorship schemes. The investors have their virtual wallets and their balances will be changed upon the commitments to invest to any student.

Note that the entities must be registered to the platform prior to accessing their personalized dashboards.

### A. Architecture of the Platform

The proposed *TEduChain* platform (see Figure 2) consists of two different operational frameworks, a relational database driven traditional framework and a distributed ledger based blockchain framework.

- *Traditional Framework:* Regular operations, such as creating user account, resetting password, sending email notifications to the users, uploading profiles, modifying personal information, resolving conflicts among the entities, providing the dashboards, and other housekeeping functionalities are managed and maintained by the traditional framework. This framework is driven by a central relational database management system (RDBMS). The traditional database stores student verification data, list of fundraisers and investors in the system and is also used for registration of the students and the investors while the blockchain database stores the students and investors details and the amount of fund given by each investor after mining is completed, ie, after a student gets the full amount needed for any particular program. The three different entities interact with the platform through their respective dashboards. Through their personalized dashboards, the students submit their applications, update their personal information, upload additional supporting documents for their application, and monitor the status of their application. Through their personalized dashboards, the fundraisers publish the list of the eligible students seeking for funds and also push the list to their pool of sponsors. The fundraisers can monitor the current status of accumulated funds of every students in the list. The sponsors can interact with the platform through their personalized dashboard. They can apply to join a new fundraiser's pool, can track the status of the students they are bidding for, can see the balance of their wallets, can add more virtual money to their wallets, and can change their terms and conditions of funding for future investments.

- **TEduChain Framework:** The contract among the student and the sponsor/s are securely stored in new blocks, mined and appended them to the chain using TEduChain framework. This framework is responsible for collects all required information from the traditional framework at the time of generating the contract among the parties. The use of blockchain here is not similar to that of bitcoin or any other cryptocurrency. Our aim is to collect funds for students and to keep it as transparent as possible. This will be a public blockchain, anyone would be able to be a part of it. There can only be three type of participants in the framework – student, fundraiser and investors. The issue of consensus in our case is solved by which fundraiser is able to collect the full funds for any particular student first. If a fundraiser is able to collect the total funds for a particular student the fundraiser will be able to add the student and the investors to the blockchain. The fundraiser also has the right to link the block to the chain.

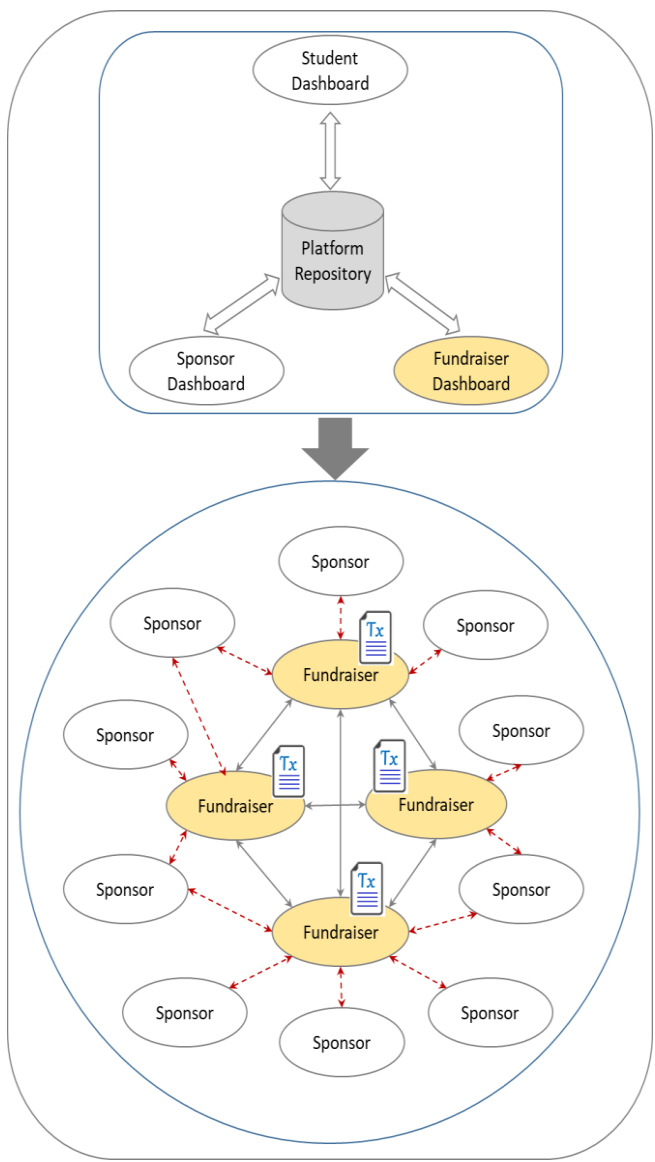

Figure 2 The architecture of the *TEduChain* platform for crowdsourcing fund for tertiary education and managing the contract afterwards.

### B. How the Platform works

Any entity interested on this platform needs to create an account first. The information required to create the account is the entity-type specific. For example, to be registered as a fundraiser, the entity should provide a valid business identification number. On the other hand, the system will require financial information in case of a sponsor.

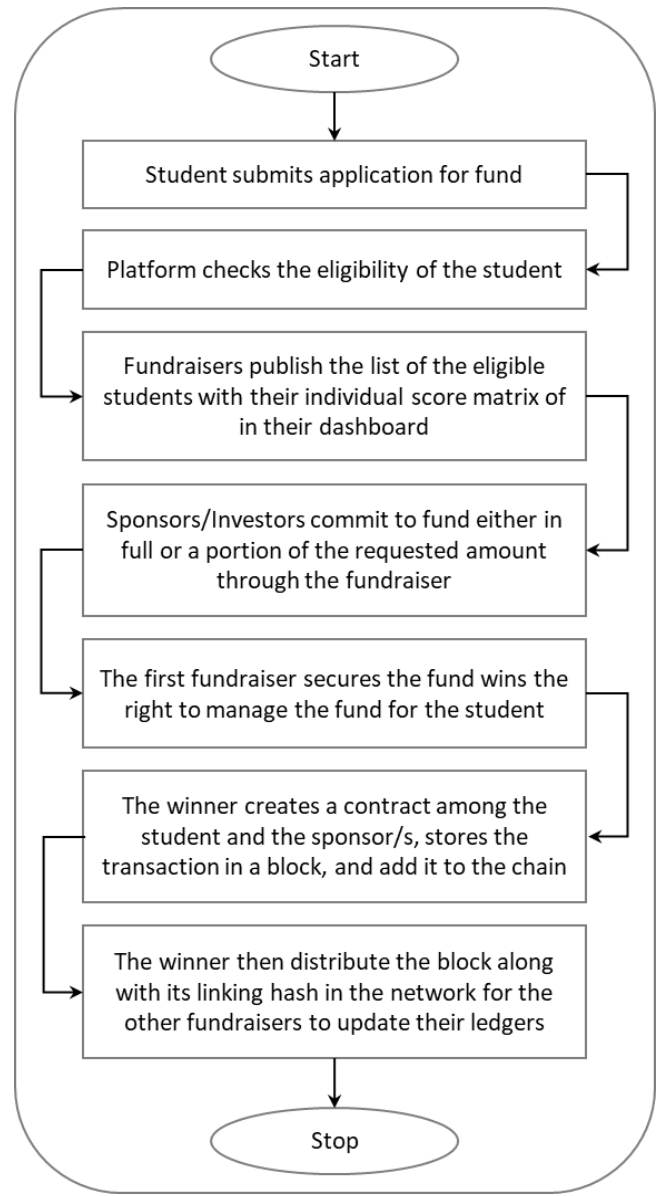

Figure 3 The workflow of the platform.

After validating the registration request by the platform, the entities will find their personalized dashboard upon sign-in to the system. They can interact with the platform using their own dashboards.

A typical workflow presented in Figure 3 begins when a student successfully sign-in to his account and landed to his personalized dashboard. The student submit a prescribed application form to the system seeking for fund. The student need to provide detail information in the form, such as name of the study program, name of the educational institute, high school results, family income, and so on.

The platform's central repository will have preloaded information to verify the eligibility of the student and notify the student with the outcome. If the student is eligible, the platform will publish the detail and the student will be automatically added to the sorted-list on the fundraisers' dashboards. The student will also be appeared in the list on the sponsors' dashboards.

A sponsor might fund a student either in full or partially through one or multiple fundraisers. The Fundraiser who will be able to secure the full amount first, will get the right to generate the contract among the student and the sponsor/s. The winner will also get the right to manage the portfolio of the student afterward. After winning the student, fundraiser will create a transaction, store that in a new block, add the block to the chain and broadcast the block and the linking hash in the blockchain network for others to update their distributed ledger. Finally, the student will be removed from the active list. The funding commitments received by the unsuccessful fundraisers from their investors are rolled back at this point and the wallet of the investors will be credited automatically.

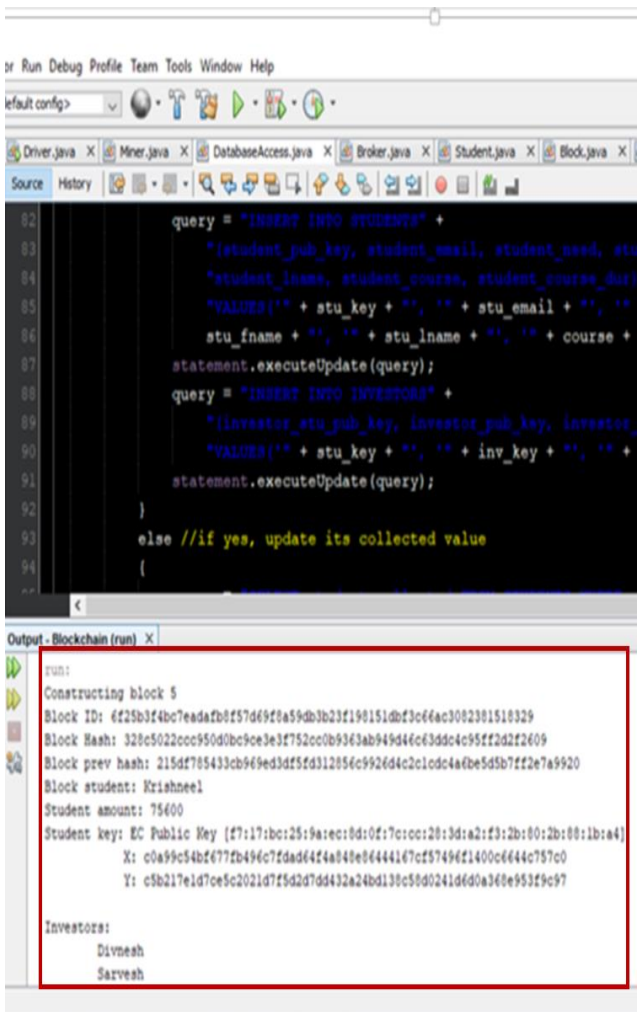

Figure 4 A snapshot of test data stored in a block during a test run of our prototype[1]

A snapshot of a block containing test data is presented in Figure 4.

The technical artifacts of blockchain implemented in TEduChain platform are presented below:

- *The mining process:* Each fundraiser manages a local database to store temporary record of active students who have received some amount by the sponsors however, yet to reach their total amount they need for a particular program to study. During the mining process, the fundraisers continuously check the balance amount in their local database that yet to be collected for every active student. Once the collected amount has reached the desired amount, the fundraiser notifies other fundraises to stop collecting fund for the student and remove him from their active list as the fundraiser has won the student. The fundraiser then goes on to prepare the block for that student and includes all the details that binds the student, the sponsor/s, and the fundraiser.

- *Linking the blocks:* Once the block is prepared, the fundraiser fetches the hash of the last block in the chain and adds it in the current block. The fundraiser then calculates the hash of the block using SHA-256 algorithm on all of the containments in the block including the hash of the last block of the chain. The fundraiser links this block to the chain and broadcast that block in the network for other fundraisers to update their distributed ledger. Once this process is completed, the fundraiser continues mining for other students.

- *Distributing the contract information:* During the linking process, the details are put into a document and sent them to the appropriate student and sponsor/s for their own reference via the email they registered with. Note that the document's hash is also stored in the block. This is just like the traditional legal document signed between the parties, however, it is generated automatically by the fundraiser after mining that includes details such as the program cost, program duration, student and respective investors, how much did each investor contribute, and benefits to the investors after the student gets a job.

- *Verifying the blockchain:* The fundraiser verifies the chain by generating the hash of the block using all of its contents excluding its own hash and comparing with its stored hash.

- *Changing block contents:* It is not possible to make changes to a previously added block however, in the case where the student or investor details need to be changed, a copy of the old block is obtained and necessary changes are made to it and added to the chain again as a new block. This does not do any modification to the old block and remains as it was. Note that only the student and investor details are allowed to be changed in the new block, such as address and contacts – the terms and conditions of the contract are immutable

---

[1] A request for the JAR file of the porotype could be made to the corresponding author.

## IV. Conclusion

Talented students should not be deprived from their tertiary study after completing high school for the lack of financial supports from their families. This platform will create opportunities for the under privileged students by outsourcing funds for their tertiary study that could not be possible by themselves.

## Acknowledgment

The authors would like to thank the anonymous reviewers for their valuable comments and suggestions to improve the quality of the paper. They are also grateful to Prof. A. Sharma, Dr. K. Chaudhary, Prof. S. Naidu, Dr. A. Sharma, and Dr. T. Sarker for their supports and guidance during the implementation and writing phase of this work.